\documentclass[superscriptaddress, twocolumn,floatfix,aps,prl,reprint]{revtex4-1}

\usepackage{graphicx}
\usepackage{dcolumn}
\usepackage[colorlinks=true]{hyperref}
\usepackage{sidecap}
\usepackage{multirow}
\usepackage{bm}
\usepackage{times}
\usepackage{color}
\usepackage{amsmath}

\newcommand{\NIII}{Na$_3$Ir$_3$O$_8$}
\newcommand{\NIV}{Na$_4$Ir$_3$O$_8$}
\newcommand{\NaI}{Na$^\text{I}$}
\newcommand{\NaII}{Na$^\text{II}$}
\newcommand{\EF}{\ensuremath{E_F}}
\newcommand{\ttg}{\ensuremath{t_{2g}}}
\newcommand{\eg}{\ensuremath{e_{g}}}
\newcommand{\jeff}{\ensuremath{J_{\text{eff}}}}
\begin{document}

\title{Fano Resonances in the Infrared Spectra of Phonons in Hyper-Kagome Na$_3$Ir$_3$O$_8$}

\author{D.~Pr\"opper}
\author{A.~N.~Yaresko}
\author{T.~I.~Larkin}
\affiliation{Max-Planck-Institut f\"ur Festk\"orperforschung, Heisenbergstra{\ss}e~1, D-70569 Stuttgart, Germany}
\author{T.~N.~Stanislavchuk}
\author{A.~A.~Sirenko}
\affiliation{Department of Physics, New Jersey Institute of Technology, Newark, New Jersey 07102, USA}
\author{T.~Takayama}
\author{A.~Matsumoto}
\author{H.~Takagi}
\affiliation{Max-Planck-Institut f\"ur Festk\"orperforschung, Heisenbergstra{\ss}e~1, D-70569 Stuttgart, Germany}
\affiliation{Department of Physics, University of Tokyo, Hongo, Tokyo 113-0033, Japan}
\author{B.~Keimer}
\author{A.~V.~Boris}
\affiliation{Max-Planck-Institut f\"ur Festk\"orperforschung, Heisenbergstra{\ss}e~1, D-70569 Stuttgart, Germany}

\date{\today}

\begin{abstract}
We report the complex dielectric function of high-quality \NIII{} single crystals determined by spectroscopic ellipsometry in the spectral range from $15\,\text{meV}$ to $2\,\text{eV}$. The far-infrared phonon spectra exhibit highly asymmetric line shapes characteristic of Fano resonances. With decreasing temperature, we observe a sharp increase of the infrared intensity of the Fano-shaped phonon modes accompanied by concomitant changes in the low energy electronic background, formed by electronic transitions between Ir 5d \ttg{} bands of a mostly $\jeff = 1/2$ character. The role of the complex Hyper-Kagome lattice structure and strong spin-orbit coupling is considered.
\end{abstract}

\maketitle
The electronic states near the Fermi energy of transition metal oxides with 5d valence electrons, compared to their 3d-electron counterparts, exhibit larger single-electron bandwidths, reduced on-site Coulomb correlations, and enhanced spin-orbit coupling. The convergence of these three energy scales has been proposed to give rise to novel electronic phases, including relativistic Mott insulators with antiferromagnetic \cite{BJKim2008,BJKim2009,JKim2012} and spin liquid \cite{Jackeli2009,Shitade2009,Choi2012,Comin2012,Gretarsson2013,Okamoto2007} ground states, for geometrically unfrustrated and frustrated lattices, respectively. Despite the prominent role of the electron-phonon interaction in the physics of 3d-electron analogues (including especially those with orbital degeneracy and geometrically frustrated lattice architectures\cite{Balents2010}), its influence on the electronic properties and phase behavior of 5d-electron compounds has not yet been addressed.

One of the most striking manifestations of the electron-phonon interaction is the quantum interference between discrete phonons and the continuum of electron-hole excitations, which leads to asymmetric Fano resonances \cite{Fano1961,Miroshnichenko2010} in the scattering and infrared (IR) absorption spectra of solids \cite{Wagner1985,Rice1977}. As the electronic contribution increases the Fano resonance may undergo a change from a phonon absorption peak to a pronounced dip corresponding to reduced absorption. This dramatic behavior is one of the characteristic properties of few-layer graphene where Fano resonances have been recently observed and studied in detail \cite{Kuzmenko2009,Tang2010,Liu2010,Li2012,Lui2013,Cappelluti2010,Cappelluti2012}. This elementary solid with tunable electron density has proved to be a model system to develop a microscopic theory which accounts quantitatively for the intensity and asymmetry of the Fano resonances \cite{Cappelluti2010,Cappelluti2012}.

In this Letter we provide evidence that conditions favorable for Fano interference are met in \NIII{}, the semi-metallic counterpart of Mott-insulating \NIV{}, one of the best candidates for a three-dimensional spin-liquid state \cite{Balents2010}. In both \NIII{} and \NIV{}, Ir ions are arranged on a three-dimensional geometrically frustrated Hyper-Kagome lattice. In formal analogy to the single-band Hubbard model \cite{BJKim2008}, one refers to the \jeff{} = 1/2 spin-orbit 5d \ttg{} states of \NIV{} as a half-filled narrow Mott-Hubbard band \cite{Lawler2008a,Lawler2008b,Zhou2008,Podolsky2009,Micklitz2010,Norman2010,Podolsky2011}, which becomes one-third filled in semi-metallic \NIII{}.  We discovered that the entire set of well-defined phonon modes in the ellipsometric IR spectra of our \NIII{} single crystals exhibits highly asymmetric line shapes characteristic of Fano resonances. We show that with decreasing temperature a sharp increase of the Fano resonances is accompanied by concomitant changes in the underlying low-energy electronic transitions. Based on fully relativistic electronic-structure calculations in the local density approximation (LDA), we attribute this unusual behavior to Rashba-type Ir 5d \ttg{} bands intersecting near the Fermi level (resembling the Dirac cone in graphene).  These bands originate from strong spin-orbit coupling combined with the broken inversion symmetry of the cubic lattice.

\begin{figure}
\includegraphics[width=7.5cm]{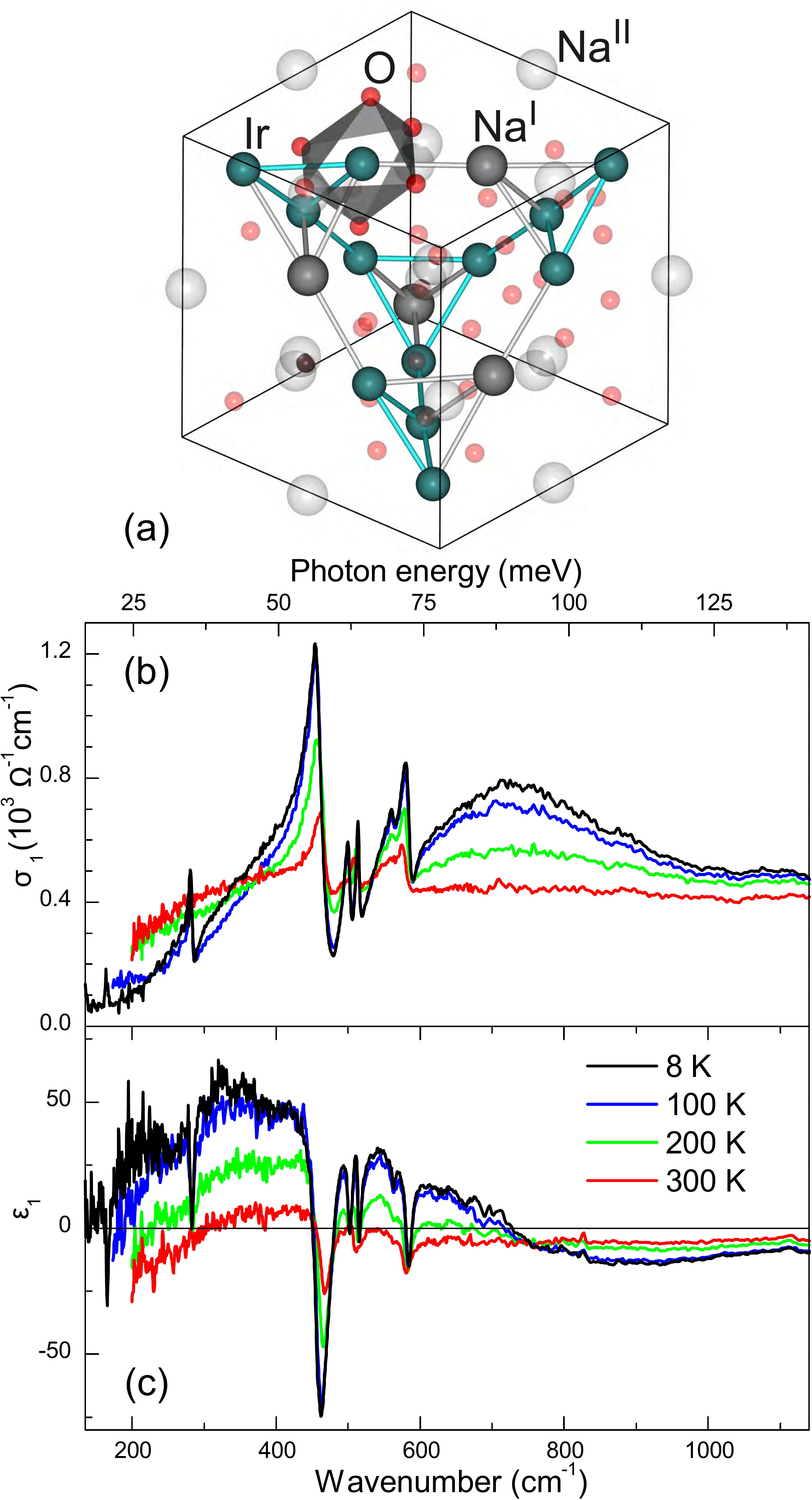}
\caption{(Color online)(a) Crystal structure of \NIII\ with the space group $P4_132$ . The viewing direction is slightly off the body diagonal $\langle 111 \rangle$ of the cubic cell. Intertwined $\text{Ir}_3\text{Na}$ tetrahedra form a Hyper-Kagome sublattice of Ir ions. Oxygen atoms occupy the corners of distorted octahedra around the Ir and \NaI{} sites. (b,c) Real part of the IR (b) conductivity $\sigma_1(\omega)$ and (c) dielectric permittivity $\varepsilon_1(\omega)$ measured at different temperatures.}
\protect\label{fig1}
\end{figure}

Single crystals \NIII\ of dimensions $0.3\times0.5\times0.5\,\text{mm}^3$ were grown from NaCl flux \cite{Takayama2013}. They crystallize in a cubic structure described by space groups $\text{P}4_132$ or $\text{P}4_332$ (No.213 and No.212 in the international tables for x-ray crystallography, respectively), with 4 formula units/unit cell [see Fig.~\ref{fig1}(a)]. X-ray diffraction analysis indicates that twelve Na atoms per unit cell sit in two fully occupied nonequivalent positions marked as \NaI{} and \NaII{} in Fig.~\ref{fig1}(a). This cubic structure can be regarded as a distorted modification of the face-centered cubic $2 \times \text{NaIr}_2\text{O}_4$ pyrochlore structure with space group $\text{Fd}\bar{3}\text{m}$, where a \NaI{} ion replaces one of the four Ir sites in the three-dimensional pyrochlore corner-sharing tetrahedron lattice $2 \times \text{Na}(\text{Na}_{1/2}\text{Ir}_{3/2})\text{O}_4$.

We report a spectroscopic ellipsometry study of \NIII{} over a wide range of temperatures (8 to 300\,K) and photon energies extending from the far infrared (IR) into the visible, from 15\,meV to 2\,eV. In the frequency range 15\,meV to 1\,eV we used home-built ellipsometers in combination with Bruker IFS 66v/S and Vertex 80v FT-IR spectrometers. Some of the experiments were performed at the IR1 beam line of the ANKA synchrotron light source at Karlsruhe Institute of Technology, Germany, and at the U4IR beam line of the NSLS synchrotron in Brookhaven National Laboratory, U.S.A \cite{Stanislavchuk2013}. The measurements in the frequency range from 0.6 to 6.5\,eV were performed with a  Woollam variable angle ellipsometer (VASE) of rotating-analyzer type. Since the dielectric response is isotropic, the complex dielectric function, $\tilde \varepsilon(\omega) = \varepsilon_1(\omega)+\mathrm{i}\varepsilon_2(\omega)$, and the related optical conductivity $\sigma_1(\omega)=\omega \varepsilon_2(\omega)/(4\pi)$, were determined by direct inversion of the Fresnel equations from the ellipsometric parameters $\Psi(\omega)$ and $\Delta(\omega)$.

Figures \ref{fig1}(b) and \ref{fig1}(c) show variation with temperature of the real part of the optical conductivity $\sigma_1(\omega)$ and permittivity $\varepsilon_1(\omega)$, respectively, in the spectral range from 120 to 1200\,$\text{cm}^{-1}$ (15 to 150\,meV). In the low-temperature spectrum one can identify a set of seven highly asymmetric phonon modes located at 166, 281, 461, 501, 514, 562, and 583\,cm$^{-1}$.
Following the Wyckoff positions of the 56 atoms and their site symmetries in the $\text{P}4_132$ unit cell of \NIII\, one would expect 21 triply degenerated $\text{T}_1$ IR-active modes \cite{Kroumova2003}. Only seven of these modes are observed in the spectra. This might be due to insufficient strength and/or weak splitting of the four $\text{T}_{1u}$ modes in the  assumptive parent $\text{Fd}\bar{3}\text{m}$ pyrochlore structure of $\text{NaIr}_2\text{O}_4$. The observed asymmetric modes are superimposed on a hump-like absorption band centered around 728\,cm$^{-1}$ (90\,meV), which grows monotonously in intensity on an almost flat background upon cooling. The downturn in $\varepsilon_1$ below 360\,cm$^{-1}$ (45\,meV) indicates the free charge carrier response, which is weak and narrow. We describe it consistently as a single Drude peak with the plasma frequency  $\omega_{\text{pl}} \approx 0.2\,\text{eV}$, in agreement with results of dc resistivity measurements \cite{Takayama2013}. The corresponding effective charge carrier density $N^\text{eff}=\frac{2m}{\pi e^2 N_\text{Ir}} \frac{\omega^2_\text{pl}}{8}\approx 0.002$ electrons per Ir atom, where $m$ is the free electron mass and $N_\text{Ir}= 1.65\times 10^{22}$cm$^{-3}$ is the density of Ir atoms. The metallic behavior of the temperature dependent dc resistivity is consistent with the narrowing of the Drude-peak at low $T$ where its spectral weight (SW) is removed from the tail and transferred to the head near the origin. As a consequence, $\sigma _{1}(\omega )$  in Fig.~\ref{fig1}(b) decreases  below 360\,cm$^{-1}$ (45\,meV) with decreasing temperature.

\begin{figure*}
\includegraphics[width=15.5cm]{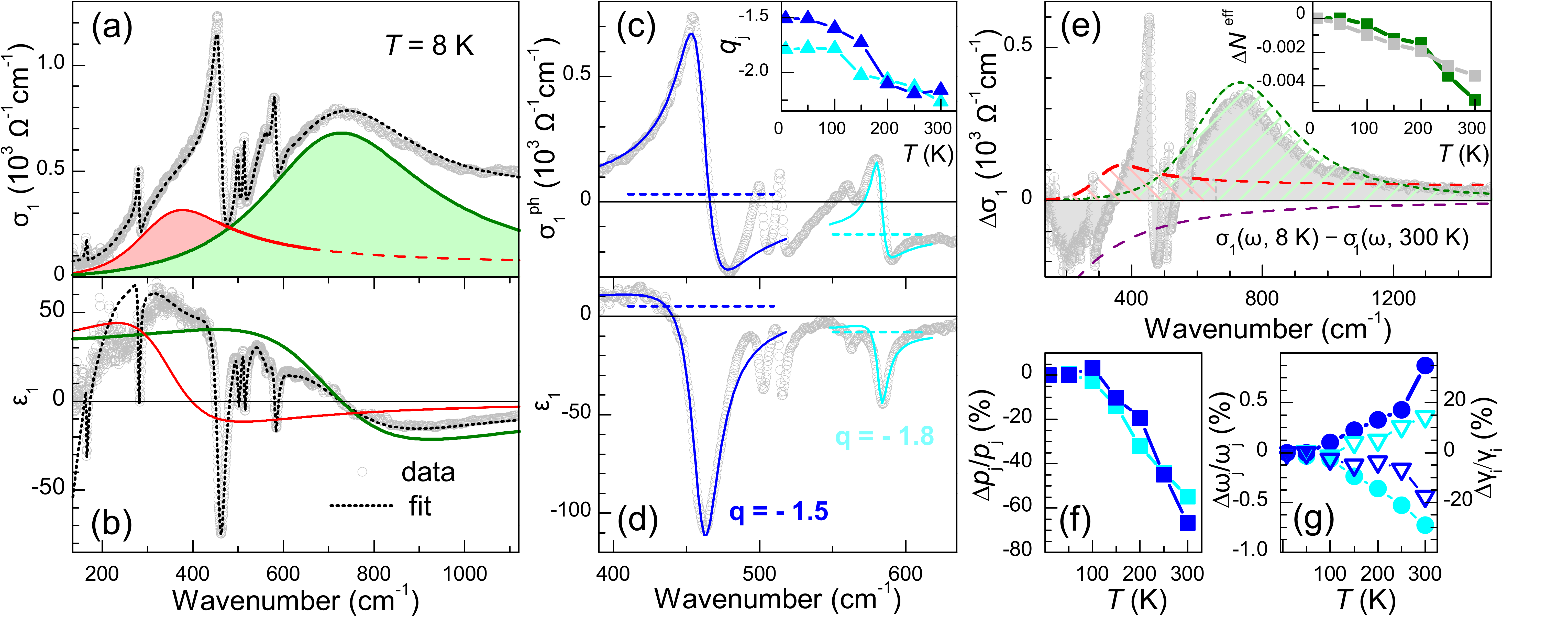}
\caption{(Color online) Real part of the IR (a) conductivity $\sigma_1(\omega)$ and (b) dielectric permittivity $\varepsilon_1(\omega)$ measured at $T=8$ K (open circles) and represented by the fit to the coupled oscillator model with Eq.(1) (dashed lines) and by the partial contribution of the low-energy electronic transitions [(a) shaded areas, (b) solid lines]. (c) and (d) Examples of the Fano fit to the spectra in the region of the  phonon modes at 461\,cm$^{-1}$ (dark blue) and 583\,cm$^{-1}$ (light cyan), obtained by subtraction of an electronic background from the data in (a) and (b). Inset: Temperature dependence of the corresponding Fano coupling parameters $q_j$. (e) Difference spectra $\sigma_1(8\,\text{K},\omega)-\sigma_1(300\,\text{K},\omega)$ (open circles) represented by the changes in the separate absorption bands (dashed lines). Inset: The SW difference with increasing temperature, integrated up to 0.5\,eV (light squares), along with the changes in the strength of the optical band at 728\,cm$^{-1}$ (90\,meV) (dark squares) as derived from the fits. (f) and (g) Relative changes in the Fano parameters with increasing temperature: (f) strength $p_j$ and (g) resonance frequency $\omega_j$ (solid circles) and width $\gamma_j$ (open triangles) for the representative modes at 461\,cm$^{-1}$ (dark blue) and 583\,cm$^{-1}$ (light cyan).
}
\protect\label{fig2}
\end{figure*}

In order to quantitatively parameterize the observed highly asymmetric far-IR modes and the strongly temperature dependent electronic background, we fit a set of generalized Lorentzian oscillators \cite{Humlicek2000} simultaneously to $\varepsilon_1(\omega)$ and $\varepsilon_{2}(\omega)$:
\begin{equation}\label{eq:GLorentz}
   \varepsilon_1(\omega)+\mathrm{i}\varepsilon_2(\omega)=\varepsilon_\infty+\sum^{K}_{j=1}\frac{\Delta\varepsilon_j\left( \omega^2_{j}-\mathrm{i}\omega\beta_j\right)}{\omega^2_{j}-\omega^2-\mathrm{i}\omega\gamma_j}
\end{equation}
with $\omega_j$, $\gamma_j$ and $\beta_j$ the resonance frequency, line width,  and asymmetry parameter, respectively, having dimensions of frequency and  dimensionless $\Delta\varepsilon_j$ being the contribution to the static permittivity. The interband transitions at high energies contribute to the background term $\varepsilon_\infty$. The causality of the dielectric response and Kramers\textendash{}Kronig requirements over the entire spectral range are maintained by holding the condition of
\begin{equation}\label{eq:sum}
\sum_{j} \Delta\varepsilon_j \beta_j=0\,.
\end{equation}
Figures~\ref{fig2}(a), \ref{fig2}(b) and Table \ref{tab:param} summarize the results of our dispersion analysis of the complex dielectric function measured at 8\,K. We unambiguously identify seven phonon modes which are highly asymmetric as indicated by the large, negative $\beta^{(\text{ph})}_j$.  In order to maintain the condition of Eq.~(\ref{eq:sum}) we used the same model with non-zero parameter $\beta^{(\text{el})}_j$ in Eq.~(\ref{eq:GLorentz}) for one of the low-energy electronic bands, but with the opposite sign compensating the high energy contributions of the phonon modes. Since the SW of broad electronic transitions is large compared to the one of the phonon modes, the asymmetry parameter $\beta^{(\text{el})}_j>0$ is relatively small and cannot be unequivocally assigned to one or another of the constituent optical bands of the electronic background. In the representative fit shown in Figs. \ref{fig2}(a) and \ref{fig2}(b) the optical band at 358\,cm$^{-1}$ (44\,meV) was selected to be asymmetric with $\beta^{(el)}_1=97$, whereas the band at 728\,cm$^{-1}$ (90\,meV) was taken to be a classical Lorentz oscillator with $\beta^{(el)}_2=0$. Some uncertainty of the phenomenological fitting procedure in the determination of the asymmetry parameter $\beta^{(el)}_j$ does not affect the parameters of the phonon modes listed in Table~\ref{tab:param}. In the vicinity of the resonance frequency the generalized Lorentz oscillator in Eq.~(\ref{eq:GLorentz}) can be converted into the Fano profile resulting from the quantum interference of a discrete state with a continuum:
\begin{equation}\label{eq:Fano}
\Delta\sigma^j_1(\omega)=\sigma^j_0\frac{(q_j+\epsilon)^2}{1+\epsilon^2}\,,
\end{equation}
where $\epsilon=2(\omega-\omega_j)/\gamma_j$ and $(2\omega_j q_j)/(q_j^2-1)\approx\beta_j$ \cite{Humlicek2000}.
Figure~\ref{fig2}(c) and \ref{fig2}(d) illustrate exemplarily the Fano profiles for the two strongest resonances at $T= 8$\,K. Both resonances are characterized by $q\approx -1.5$ with an antisymmetric absorption line shape implying comparable phonon and electronic contributions into the Fano resonances. The temperature evolution of $q_j$ is shown in the inset of Fig.~\ref{fig2}(c) and one can see that the electron-phonon coupling weakens with increasing temperature.

\begin{table}
\caption{\label{tab:param} Best fit results for the Fano resonances and low energy electronic transitions in the far IR spectral range of \NIII\ at $T=8\,\text{K}$ with $\omega_j$, $\Delta\varepsilon_j$, $\gamma_j$, $\beta_j$ and $q_j$  being the contribution to the static permittivity, resonance frequency, line width and asymmetry parameters, respectively.}
\begin{ruledtabular}
\begin{tabular}{l l r  r  r  r r}
   & j & \multicolumn{1}{c}{$\omega_j$} & \multicolumn{1}{c}{$\Delta\varepsilon_j$}  &   \multicolumn{1}{c}{$\gamma_j$} & \multicolumn{1}{c}{$\beta_j$} & \multicolumn{1}{c}{$q_j$}\\
    & & \multicolumn{1}{c}{$(cm^{-1})$}  & & \multicolumn{1}{c}{$(cm^{-1})$} & \multicolumn{1}{c}{$(cm^{-1})$}\\
    \hline
(ph) &1 & 166 & 0.70  &  3.0 & -101     & -3.6\\
& 2 & 281 & 0.56  &  5.0 & -535     & -1.7\\
   &  3 & 461 & 2.33  &  22.6 & -1092 & -1.5\\
    & 4 & 501 & 0.33  &  6.8 & -490     & -2.5\\
    & 5 & 514 & 0.27  &  4.3 & -318     & -3.5\\
    & 6 & 562 & 0.05  &  7.3 & -400     & -3.1\\
     & 7 & 583 & 0.31  &  8.9 & -950     & -1.8\\
\hline
 (el) & 1 & 358 & 36.1  &  249 & 97     & 7.4\\
 & 2 & 728 & 34.4  &  447 &      & \\
\end{tabular}
\end{ruledtabular}
\end{table}

Figure~\ref{fig2}(e) displays the changes occurring in the IR conductivity spectra with increasing temperature, as shown by the difference spectra $\Delta\sigma_1(\omega)=\sigma_1(\omega, 8\,\text{K})-\sigma_1(\omega, 300\,\text{K})$. By using the dispersion analysis of the dielectric function spectra at different temperatures with Eq.~(\ref{eq:GLorentz}), we determine more accurately the temperature variation of the constituent Drude and low-energy inter-band contributions shown by the dashed lines in Fig.~\ref{fig2}(e). The corresponding loss of the associated SW, $\text{SW}_j=\frac{\pi}{120}\Delta\varepsilon_j\omega_j^2$, with increasing temperature above 8\,K is detailed in the inset of Fig.~\ref{fig2}(e) for the optical band at 728\,cm$^{-1}$ (90\,meV), along with the temperature dependence of the integrated SW difference, $\Delta\text{SW}(T)=\int_0^{1 eV} [\sigma_1(\omega,T)-\sigma_1(\omega,8\,\text{K})] \text{d}\omega$. In the latter case, the SW in the extrapolation region below 15\,meV was determined by the procedure based on the Kramers\textendash{}Kronig consistency check of the experimentally measured $\Delta\varepsilon_1(\omega)$ and $\Delta \sigma_1(\omega)$, $15\,\text{meV}<\omega<1\,\text{eV}$ (See Methods in Ref.~\cite{Charnukha2011}). The temperature-driven loss of the SW, expressed in terms of the effective number of charge carriers $\Delta N^\text{eff}(T)=\frac{2m }{\pi e^2 N_\text{Ir}}\Delta\text{SW}$, is comparable to the free carrier density evaluated above, and it is not recovered within a high energy scale in excess of 1 eV. In order to quantify the temperature-dependent intensity of the sharp Fano resonances we refer to the strength parameter $p_j$  \cite{Cappelluti2010}, which accounts for its both positive and negative contributions to $\Delta\sigma^j_1(\omega)$ (approximately equal for $q_j\approx1$): $p_j=\Delta\varepsilon_j \omega_j^2(1+\frac{1}{q_j^2})/8$. Figure~\ref{fig2}(f) shows the temperature evolution of the Fano-strength parameter $p_j$ for the two representative resonances. It appears clear from our analysis that both the intensity and Fano-asymmetry variations of the sharp resonances correlate remarkably well with the strength of the underlying electronic transitions. Therefore the SW loss of the phonon resonances stems from the SW loss of the interband transition. The observed intensity and asymmetry (Table~\ref{tab:param}) of the Fano resonances imply a considerable electron-phonon coupling in \NIII{}, given the high bonding ionicity and the large number of IR active modes with evidently large intrinsic dipole moments. This is in stark contrast to the case of graphene, where the static intrinsic dipole is negligible and similar values of the Fano parameter $q$ \cite{Kuzmenko2009,Tang2010,Liu2010,Li2012,Lui2013} originate mainly from the electronic excitations in the presence of a relatively small electron-phonon coupling \cite{Cappelluti2010}. Further evidence for the strong interaction of the phonons with the underlying electronic transitions in \NIII{} is the temperature dependence of the resonance frequency and width of the strongest mode at 461\,cm$^{-1}$ (dark solid circles and open triangles in Fig.~\ref{fig2}(g), respectively), which hardens and narrows with increasing temperature. A quantitative estimate of the electron-phonon coupling constants and the Born effective charges requires a detailed analysis of the lattice dynamics in this complex Hyper-Kagome structure, which is an important challenge for future theoretical work.

\begin{SCfigure*}
\centering
\includegraphics[width=0.63\textwidth]{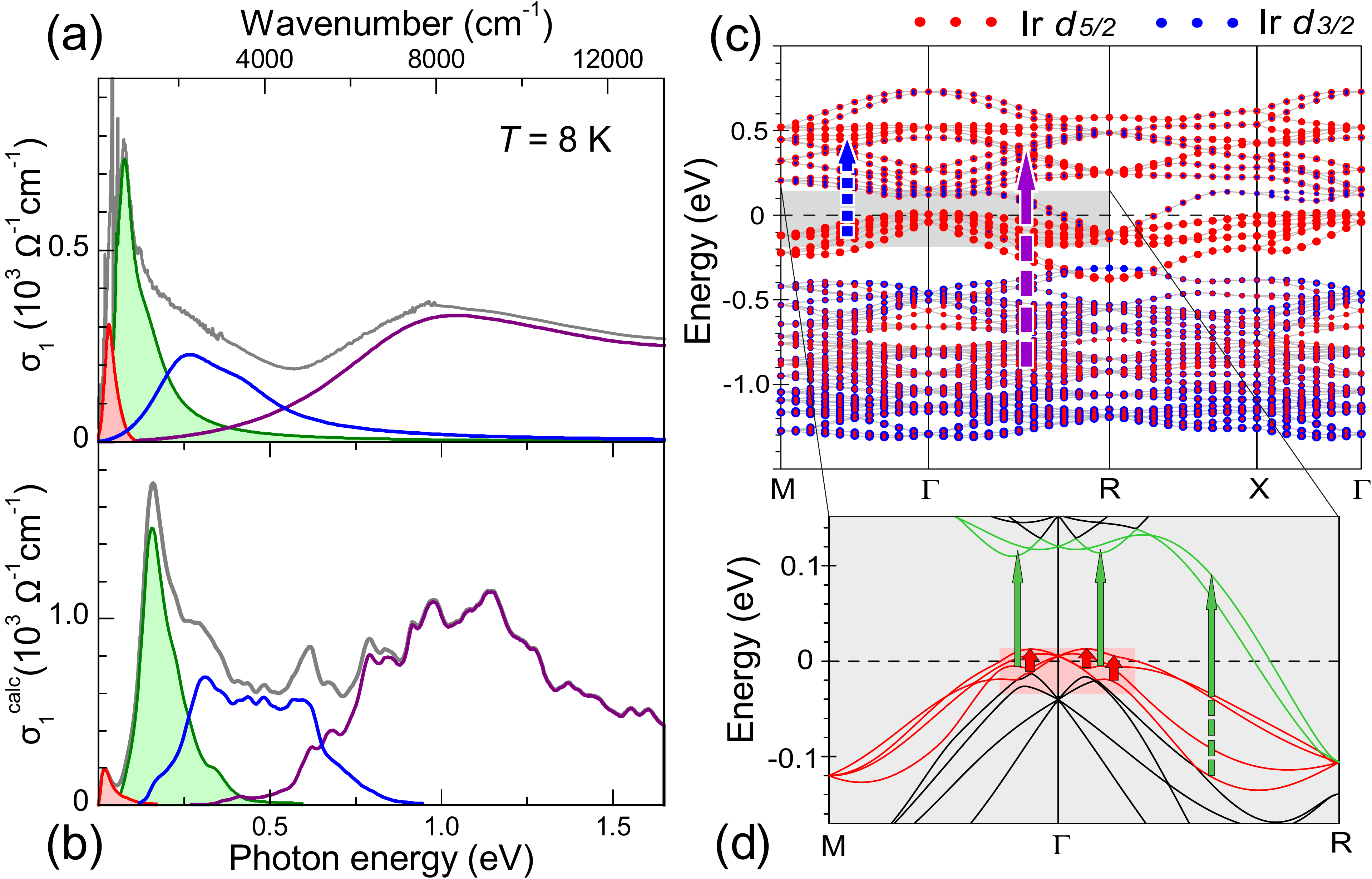}
\caption{(Color) (a) Real part of the optical conductivity of \NIII{} measured at $T=8$\,K and contributing interband transitions determined by a dispersion analysis. (b) Corresponding relativistic LDA calculation with a breakdown into separate \ttg{} orbital contributions, as described in the text. (c) Band structure for Ir \ttg{} states from the same LDA calculation. The size of blue and red circles is proportional to the weight of $d$ states with true $d_{3/2}$ and $d_{5/2}$ orbital character, respectively. (d) Enlargement of the band structure of (c) near the Fermi energy. Color coding of the allowed optical transitions sketched with arrows in (c) and (d) corresponds to the band color in (a) and (b).
}
\protect\label{fig3}
\end{SCfigure*}

In the present study, we take the first step in this direction by identifying the origin of the electronic bands coupled to the phonons. To this end, we perform relativistic LDA band structure calculations using the linear muffin-tin orbital (LMTO) method \cite{Andersen1975} for the experimental crystal structure of \NIII{} \cite{Takayama2013}. The details of the implementation of the fully relativistic LMTO method and calculations of the optical spectra can be found in Ref.~\cite{book:AHY04}. Although the O$_6$ octahedra surrounding the Ir ions in \NIII{} are distorted, the Ir $d$ shell is split into well separated \textquotedblleft{}\ttg\textquotedblright{} and \textquotedblleft{}\eg\textquotedblright{} states. A simple electron count shows that the \ttg{} states of each Ir are filled by 4.67 electrons. In spite of the noninteger formal valency of an Ir$^{4.33+}$ ion scalar-relativistic calculations which neglect spin-orbit coupling (SOC) give an insulating solution with 8 unoccupied doubly degenerate bands separated by a gap of $\sim\,0.2$\,eV from the remaining occupied \ttg{} bands. Fully relativistic Ir \ttg{} bands decorated with circles proportional to the weight of Ir $d_{3/2}$ and $d_{5/2}$ states are shown in Fig.~\ref{fig3}(c).
Since the crystal structure lacks inversion symmetry the Kramers degeneracy is lifted everywhere except for time-reversal invariant points. A striking effect of strong SOC within the Ir $d$-shell is that a pair of unoccupied \ttg{} bands closes the gap and creates two large electron-like Fermi surfaces around the $R$ point. The charge balance is then maintained by depopulating hole-like bands near $\Gamma$ rendering the material semi-metallic, with a negative indirect band gap between $\Gamma$ and $R$. In semi-metallic materials strong temperature effects in the optical spectra are expected \cite{Menzel2009}. As temperature increases the phonon assistance of the indirect transitions is dramatically enhanced, which changes the valence and conduction band occupancies at the $R$ and $\Gamma$ points, respectively. This may explain the observed temperature evolution of the low energy electronic background with its spectral weight redistribution over an extended energy range.

The relativistic \ttg{} bands are subdivided into a lower- and a higher-energy group of 48 and 24 bands, respectively, reflecting the splitting of Ir \ttg{} states into a quartet with $\jeff=3/2$ and a doublet with $\jeff=1/2$ by strong SOC. As the functions of the $\jeff=1/2$ doublet are given by linear combinations of $d_{5/2}$ states only. Consequently, the appreciable contribution of $d_{3/2}$ states to unoccupied \ttg{} bands [Fig.~\ref{fig3}(c)] implies a strong hybridization between the $\jeff=3/2$ and $\jeff=1/2$ states. An expanded view of the relativistic bands near \EF{} is shown in Fig.~\ref{fig3}(d) where the two bands which close the scalar-relativistic gap are plotted by green lines while four $\jeff=1/2$ bands which become partially occupied in order to keep the charge neutrality are shown in red. These transitions are responsible for a peak of the optical conductivity below 0.1 eV shaded by red in Fig.~\ref{fig3}(b). The bands crossing \EF{} also provide an intra-band Drude contribution to the conductivity with $\omega_\text{pl}=0.6$\,eV (not shown in Fig.~\ref{fig3}(b)) which is somewhat larger than the experimentally observed value.  A sharp peak at 0.2 eV (green shaded area) is due to interband transitions from occupied $\jeff=1/2$ bands to the pair of bands which cross \EF{} near $R$, whereas transitions from the same initial $\jeff=1/2$ bands to the rest of the unoccupied \ttg{} bands give a broad feature centered at 0.5\,eV. Finally, transitions from the bands formed predominantly by $\jeff=3/2$ states to unoccupied \ttg{} bands give a broad maximum of the conductivity at $\sim$1.2\,eV.  The overall shape of the optical conductivity, i.e. a strong, low lying transition followed by a near IR band peaking at about 1\,eV and an absorption edge setting in between 2 and 3\,eV, is nicely captured by our calculations, although the absolute values calculated exceed the experimental ones roughly by a factor of two.

The results presented above emphasize the significance of the electron-phonon interaction in strongly spin-orbit driven systems. Because of the lack of inversion symmetry the four $\jeff=1/2$ bands in \NIII{} have linear Rashba-type dispersion in the vicinity of $\Gamma$ with their apex lying slightly above \EF{}, similar to the bulk band structure of BiTeI \cite{Ishizaka2011,Lee2011}. An analysis of the dipole matrix elements has shown that interband transitions between these partially filled bands have a high probability. This provides a high density of electron-hole excitations which interfere with superimposed discrete phonon states, in a similar manner as discussed for graphene \cite{Kuzmenko2009,Tang2010}. The geometrical frustration on the Hyper-Kagome lattice may facilitate the electron-phonon coupling effects. Our experiments highlight the need for a detailed analysis of the electron-phonon interaction and its influence on the ground states and low energy excitation spectra in the 5d-electron compounds.

\section{acknowledgements}
The project was supported by the German Science Foundation under Grant No. BO 3537/1-1 within SPP 1458. We gratefully acknowledge Y.-L.~Mathis for support at the IR beam line of the synchrotron facility ANKA at the Karlsruhe Institute of Technology, {G.~L.~Carr} for support at the U4IR beam line of the NSLS synchrotron further enhance at the Brookhaven National Laboratory. The NSLS is operated as a User Facility for the U.S. Department of Energy under Contract No. DE-AC02-98CH10886.

\end{document}